\documentclass[reprint,aps,pra,groupedaddress,superscriptaddress,amsmath,amssymb,showpacs]{revtex4-1}

\usepackage{graphicx}
\usepackage{dcolumn}
\usepackage{bm}

\usepackage[mathlines]{lineno}
\usepackage{amssymb,epsfig}
\usepackage{color}

\begin{document}

\title{Ionic vibration induced transparency and Autler-Townes splitting}

\author{Wenjun Shao}
\affiliation{Department of Physics, Shanghai Normal University, Shanghai
200234, China}

\author{Fei Wang}
\affiliation{College of Science, China Three
Gorges University, Yichang 443002, China}
\affiliation{Centre for Quantum Technologies, National University of Singapore, 2 Science
Drive 3, Singapore 117542} 

\author{Xun-Li Feng}
\email{xlfeng@shnu.edu.cn}
\affiliation{Department of Physics, Shanghai Normal University, Shanghai
200234, China}

\author{C. H. Oh}
\affiliation{Centre for Quantum Technologies, National University of Singapore, 2 Science
Drive 3, Singapore 117542} 
\affiliation{Department of Physics, National University of Singapore, 2 Science Drive 3,
Singapore 117542}

\begin{abstract}
In this work, the absorption spectrum of a two-level ion in a linear Paul
trap is investigated, the ion is supposed to be driven by two orthogonal
laser beams, the one along the axial of the trap acts as the control light
beam, the other as probe beam. When the frequency of the control laser is
tuned to the first red sideband of the ionic transition, the coupling
between the internal states of the ion and vibrational mode turns out to be
a Jaynes-Cummings (JC) Hamiltonian, which together with the coupling between
the probe beam and the two-level ion constructs a $\Lambda$-type three-level
structure. In this case the transparency window may appear in the absorption
spectrum of the probe light, which is induced by the ionic vibration and is
very similar to the cavity induced transparency [1996 {\it Opt. Commun.} {\bf 126} 
230-235]. On the other hand, when the frequency of the control laser
is tuned to the first blue sideband of the ionic transition, the two-level
ion and vibrational mode are governed by an anti-Jaynes-Cummings (anti-JC)
Hamiltonian, the total system including the probe beam forms a $V$-type
three-level structure. And the Autler-Townes splitting in the absorption
spectrum is found.
\end{abstract}

%\pacs{42.50.Gy, 37.10.Ty}
\maketitle

\section{Introduction}

Quantum interferences, which may occur in many quantum processes along
alternative pathways, play a very significant role in quantum mechanics. The
superpositions of the probability amplitudes in different pathways give rise
to phenomena analogous to constructive and destructive interference between
classical waves. In quantum optics many valuable applications of quantum
interferences such as coherent population trapping \cite{CPT,CPT2}, lasing
without inversion \cite{LWI,LWI2} and electromagnetically induced
transparency (EIT) \cite{EIT1,EIT2,EIT3} have been examined. As for the EIT,
it is the quantum interference that makes the transparency of a weak probe
light through an originally opaque atomic medium in a narrow spectral window
with the help of a strong control laser beam. Electromagnetically induced
transparency has been studied extensively and generalized in different ways;
several EIT-like phenomena have also been predicted and some of them have
been observed experimentally. For instance, Rice and Brecha predicted that,
in cavity QED where a single-mode cavity contains a two-level atom, a hole
in the absorption spectrum of the atom may emerge at line center for a weak
probe light, this phenomenon is exactly due to the quantum interference
between two different transition paths induced by the cavity field, and it
is called cavity induced transparency (CIT) \cite{CIT}. It was shown that
the vacuum Rabi splitting in a cavity QED can even induce the transparency
of a probe light in the $\Lambda$-type three-level system, called vaccum
induced transparency \cite{VaIT0,VaIT}. In a cavity-optomechanical system 
\cite{COM}, the optomechanically induced transparency was realized in
experiment \cite{OMIT1,OMIT2}, in which an optical light, tuned to a
sideband transition of a micro-optomechanical system, acts as control field,
and the intracavity field acts as a probe field, such a system forms a $%
\Lambda $-type three-level structure, and the presence of the control field
can thus induce the transparency of the probe field. Electromagnetically
induced transparency and EIT-like phenomena have many potential applications
in controlling the propagation properties of the medium for light such as
absorption coefficient, refractive index, propagating speed and nonlinearity
etc. \cite{Lukin}, and in quantum information processing such as quantum
information memory \cite{Polariton}.

Besides, a phenomenon similar to EIT, known as Autler-Townes splitting (ATS) 
\cite{ATS1, ATS2, ATS3}, also displays a dip in the absorption spectrum of a
weak probe field in a quantum system appropriately coupling to a strong
driving field. Differently, ATS is not contributed to destructive
interference but to the driving-field-induced shift of the transition
frequency \cite{EITATS}. Abi-Salloum analyzed three-level systems: $\Lambda$%
, $V$ and two ladder with upper- and lower-level driving respectively, and
found EIT mainly appears in $\Lambda$ and upper-driven ladder three-level
systems and ATS in V and lower-driven ladder three-level systems \cite%
{EITATS}. Anisimov et al. proposed an objective method on discerning ATS
from EIT \cite{EA1}.

On the other hand, ion traps have been developed to be a state-of-the-art
technique for solving problems in quantum mechanics, quantum optics and
quantum information processing etc., for instance, the famous
Jaynes-Commings (JC) model and various generalized JC models were realized
experimentally for two-level ions and the phonons of the ionic vibration 
\cite{RMP-ion}; the controlled-NOT operation proposed by Cirac and Zoller 
\cite{Cirac-Zoller} was realized in experiment as well \cite{Cirac-ZollerEXP}%
. Moreover, very recently, high-fidelity trapped-ion-based quantum logic
gates \cite{iontrapEXP,Wineland2} and those with multi-element qubits \cite%
{Wineland1} were demonstrated; and Shor's algorithm \cite{Blatt1} and the
quantum simulation of lattice gauge theories \cite{Blatt2} were realized as
the first step towards a real quantum computer.

As trapped ions can provide us a system governed by JC or anti-JC
Hamiltonian as cavity QED system does, one may naturally ask the following
question whether one can resort to the ionic vibration to realize the
transparency of a probe light. In this paper, we give a positive answer to
this question. Our results show when the control laser light is tuned to the
first red sideband of the ionic transition (corresponding to the JC model),
a transparency window in the absorption spectrum of the probe light emerges,
we refer to such a phenomenon as ionic vibration induced transparency (VIT).
And when the control laser light is tuned to the first blue sideband of the
ionic transition (corresponding to the anti-JC model), Autler-Townes
splitting emerges, which also displays a dip (or reduction, hole) in the
absorption spectrum.

The rest of this paper is organized as follows, in Sec. 2, we describe the
theoretical model of our scheme and the Hamiltonian for the driven trapped
ion in Lamb-Dicke regime. In Sec. 3 we investigate the VIT when the
frequency of the control light is tuned to the first red sideband of the
ionic transition; in Sec. 4 we show the ATS when the frequency of the
control light is tuned to the first blue sideband of the ionic transition.
Finally, we end with discussion and conclusion in Sec. 5.

\section{Theoretical model}

The system we consider here is a single two-level ion confined in a linear
Paul trap, where the strength of the radial confinement is assumed to be
largely stronger than that along the axial direction, the movement in the
radial direction can thus be ignored \cite{RMP-ion} and one considers only
the center-of-mass mode of the ionic vibration. We assume the ion in the
Paul trap is driven by two orthogonal laser beams, one is along the axial
(or longitudinal) direction of the trap, the other along the radial (or
transverse) direction. We further assume the longitudinal laser beam is a
traveling wave with frequency $\omega _{L}$\textbf{\ }and it acts as the
control light beam. The transverse laser beam is a
weak laser light with a tunable frequency $\omega _{P}$ and it acts as the
probe field.

As the motion of the ion is mainly in the longitudinal direction and the
transverse motion can almost be ignored, the Hamiltonian for this system is
thus given by \cite{RMP-ion} 
\begin{eqnarray}
H&& =\frac{\hbar }{2}\omega _{a}\sigma _{z}+\hbar (\nu a^{\dagger }a+\frac{1%
}{2}) +\frac{\hbar }{2}\Omega \left( \sigma _{+}+\sigma _{-}\right) \times \nonumber \\
&& \left[
e^{i\eta \left( a^{\dagger }+a\right) -i\omega _{L}t}+e^{-i\eta \left(
a^{\dagger }+a\right) +i\omega _{L}t}\right]\nonumber \\
&&
+ i\hbar \varepsilon \left( \sigma _{-}e^{i\omega
_{P}t}-\sigma_{+}e^{-i\omega _{P}t}\right)
,  \label{1}
\end{eqnarray}
where $\sigma _{z}=\left\vert e\right\rangle \left\langle e\right\vert -$ $%
\left\vert g\right\rangle \left\langle g\right\vert ,$ $\sigma
_{+}=\left\vert e\right\rangle \left\langle g\right\vert ,$ and $\sigma
_{-}=\left\vert g\right\rangle \left\langle e\right\vert $ with $\left\vert
e\right\rangle $ and $\left\vert g\right\rangle $ being the excited and
ground states of the ion respectively, $a^{\dagger }$ and $a$ are the
creation and annihilation operators for the center-of-mass motion of the
trapped ion respectively, $\Omega $ ( $\varepsilon $\ ) is the Rabi
frequency of the longitudinal (transverse) laser field, and $\eta =k_{L}/%
\sqrt{2m\nu }$ is the Lamb-Dicke parameter, with $k_{L}=\omega _{L}/c$ being
the wave vector of the longitudinal laser field.

We suppose the trapped ion is constrained in the Lamb-Dicke regime, and the
Lamb-Dicke parameter $\eta $ meets the condition $\eta \ll 1$. The
Hamiltonian $H$\ can be approximated by the expansion to the first order in $%
\eta $, 
\begin{eqnarray}
H && \approx \frac{\hbar }{2}\omega _{a}\sigma _{z}+\hbar (\nu a^{\dagger }a+%
\frac{1}{2}) +\frac{\hbar }{2}\Omega \left( \sigma _{+}+\sigma _{-}\right) \times  \nonumber \\
&& \left[
e^{-i\omega _{L}t}\left( 1+i\eta a+i\eta a^{\dagger }\right) +e^{i\omega
_{L}t}\left( 1-i\eta a-i\eta a^{\dagger }\right) \right]\nonumber \\
&&+i\hbar \varepsilon \left( \sigma _{-}e^{+i\omega _{P}t}-\sigma
_{+}e^{-i\omega _{P}t}\right).
\end{eqnarray}

\begin{figure}[btp]
\begin{center}
\includegraphics[
height=2.1in,
width=3.6in
]{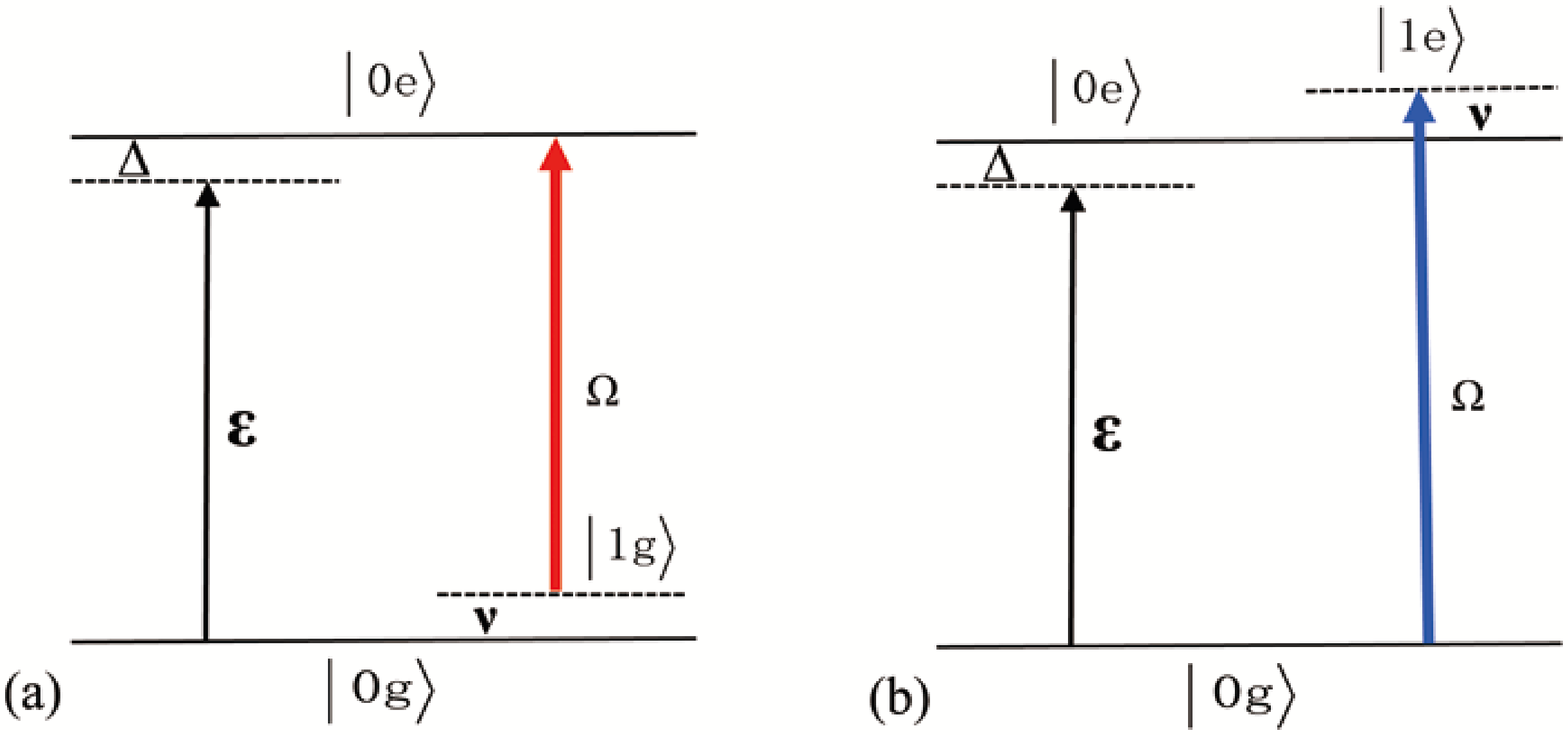}
\end{center}
\caption{Transition processes in $(a)$ red-detuning case and $(b)$
blue-detuning case.}
\end{figure}

In the following we will investigate the two different cases in which the
frequency of the longitudinal laser beam, $\omega _{L}$, is respectively
tuned to the first red sideband or the first blue sideband of the ionic
internal transition frequency $\omega _{a}$. In the red-detuning case, when
the frequency of the longitudinal laser beam is tuned to the first red
sideband of the atomic transition, the frequency of the control light
satisfies $\omega _{L}=\omega _{a}-\nu $; we apply a unitary transformation
to the Hamiltonian $H\ $to deal with the counterrotating-wave terms in Eq. ($%
2$), 
\begin{equation}
H^{\prime }=i\hbar \frac{\partial U_{R}^{\dagger}}{\partial t}
U_{R}+U_{R}^{\dagger}HU_{R},  \label{3}
\end{equation}
where 
\begin{equation}
U_{R}=\exp \left\{ -i\left[ \frac{1}{2}\omega _{P}\sigma _{z}+\left( \omega
_{P}-\omega _{L}\right) a^{\dagger }a\right] t\right\} .  \label{4}
\end{equation}
Then the Hamiltonian of the system can be simplified by discarding the
rapidly oscillating terms (taking the rotating-wave approximation), and we
finally obtain 
\begin{equation}
H^{\prime }=H_{JC}+i\hbar \varepsilon \left( \sigma _{-}-\sigma _{+}\right) ,
\label{5}
\end{equation}
where $\Delta =\omega _{a}-\omega _{P}$ is the detuning between the ionic
transition and probe field and $H_{JC}$ is the JC Hamiltonian taking the
form 
\begin{equation}
H_{JC}=\frac{\hbar }{2}\Delta \sigma _{z}+\hbar \Delta a^{\dagger }a+\frac{%
i\hbar }{2}\eta \Omega \left( \sigma _{+}a-\sigma _{-}a^{\dagger }\right) .
\label{6}
\end{equation}

In the blue-detuning case, the frequency of the longitudinal laser beam is
setted to $\omega _{L}=\omega _{a}+\nu $, that is, the longitudinal laser is
tuned to the first blue sideband of the atomic transition. Here we apply the
following unitary transformation to the Hamiltonian $H$ (Eq. ($2$)), 
\begin{equation}
H^{\prime \prime }=i\hbar \frac{\partial U_{B}^{\dagger}}{\partial t}%
U_{B}+U_{B}^{\dagger}HU_{B},  \label{7}
\end{equation}
where 
\begin{equation}
U_{B}=\exp \left\{ -i\left[ \frac{1}{2}\omega _{P}\sigma _{z}+\left( \omega
_{L}-\omega _{P}\right) a^{\dagger }a\right] t\right\} .
\end{equation}%
Similar to the red-detuning case, we obtain an anti-JC Hamiltonian by
utilizing the rotating-wave approximation 
\begin{equation}
H^{\prime \prime }=H_{AJC}+i\hbar \varepsilon \left( \sigma _{-}-\sigma
_{+}\right) ,
\end{equation}
and the anti-JC Hamiltonian takes the form 
\begin{equation}
H_{AJC}=\frac{\hbar }{2}\Delta \sigma _{z}-\hbar \Delta a^{\dagger }a+\frac{%
i\hbar }{2}\eta \Omega \left( \sigma _{+}a^{\dagger }-\sigma _{-}a\right) .
\label{10}
\end{equation}

If we concentrate on the case that the motion of the ion is nearly confined
to its ground state and the probe light is very weak, only the zero- and
one-phonon states of the vibration of the ion need to be taken into account.
Following Ref. \cite{CIT} the total states of the system in the red-detuning
case are now spanned by $\{\left\vert 0g\right\rangle ,$\ $\left\vert
0e\right\rangle ,$\ $\left\vert 1g\right\rangle \}$, where $\left\vert
0g\right\rangle \equiv \left\vert 0\right\rangle \otimes \left\vert
g\right\rangle $ and so on, here the numerical index,\textit{\ }$0$ or%
\textit{\ }$1$, indicates the phonon number state. The energy level
structure of these three states is sketched in Fig. $1(a)$. Similarly, in
the blue-detuning case the total states of the system are spanned by\textbf{%
\ }$\mathbf{\{}\left\vert 0g\right\rangle ,$\textbf{\ }$\left\vert
0e\right\rangle ,$\textbf{\ }$\left\vert 1e\right\rangle \}$. The energy
level structure in this case is sketched in Fig. $1(b)$.

\begin{figure}[btp]
\begin{center}
\includegraphics[
height=5.5in,
width=3.3in
]{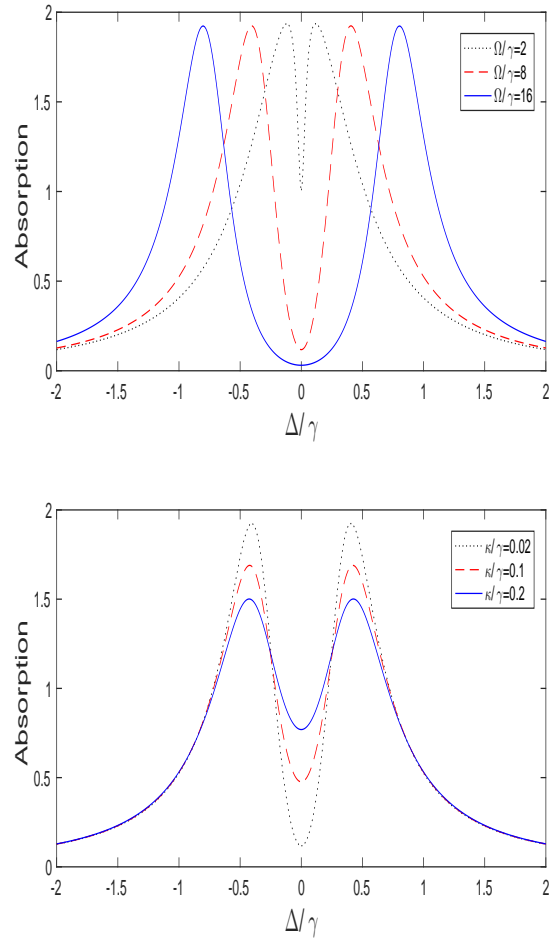}
\end{center}
\caption{The absorption spectra for the probe light when the control light
is tuned to the first red sideband. The vertical axis is $Im[\protect\rho %
_{0g;0e}/i\protect\varepsilon ]$ in the unit of $1/\protect\gamma $ and the
horizontal axis is $\Delta /\protect\gamma $ with the Lamb-Dicke parameter $%
\protect\eta =0.1$. $(a)$ $\protect\gamma /\protect\kappa =50$ and $\Omega /%
\protect\gamma =2$ (black dotted line), $\Omega /\protect\gamma =8$ (red
dashed line), $\Omega /\protect\gamma =16$ (blue solid line). $(b)$ $\Omega /%
\protect\gamma =8$ and $\protect\kappa /\protect\gamma =0.02$ (black dotted
line), $\protect\kappa /\protect\gamma =0.1$ (red dashed line), $\protect%
\kappa /\protect\gamma =0.2$ (blue solid line). }
\end{figure}

\section{VIT in red-detuning case}

Now let us consider the spontaneous emission of the ionic excited state and
the heating effect of the vibrational motion induced by coupling to the
environment. We suppose the interaction between the ionic internal states
and its reservoir is weak, so is the interaction between the vibrational
mode and its reservoir. Thus one can adopt the Born and Markov
approximations to deal with spontaneous emission and the heating effect, and
the master equation for such a system takes the form 
\begin{eqnarray}
\dot{\rho} &=&-\frac{i}{\hbar }\left[ H^{\prime },\rho \right] +\kappa (%
\overline{n}+1)\left( 2a\rho a^{\dagger }-a^{\dagger }a\rho -\rho a^{\dagger
}a\right)  \nonumber \\
&&+\kappa \overline{n}\left( 2a^{\dagger }\rho a-aa^{\dagger }\rho -\rho
aa^{\dagger }\right) \nonumber \\
&&+\frac{\gamma }{2}\left( 2\sigma _{-}\rho \sigma
_{+}-\sigma _{+}\sigma _{-}\rho -\rho \sigma _{+}\sigma _{-}\right) ,
\end{eqnarray}%
where $\kappa $ is the heating rate for vibrational motion and $\overline{n}$
is the average thermal phonon number, $\gamma $ is the spontaneous emission
rate, here we have assumed that the ionic excited state couples to the
vacuum reservoir of the electromagnetic field. As supposed above, the
vibration of the ion is nearly confined to its ground state, so the average
thermal phonon $\overline{n}$ is almost zero.

The elements of the density matrix in the states $\{\left\vert
0g\right\rangle ,$\ $\left\vert 0e\right\rangle ,$\ $\left\vert
1g\right\rangle \}$ take the following form according to the master equation
($11$):%
\begin{eqnarray}
\dot{\rho}_{0g;0g} &=&\gamma \rho _{0e;0e}+2\kappa \rho _{1g;1g}+\varepsilon
\left( \rho _{0e;0g}+\rho _{0g;0e}\right) ,  \label{12} \\
\dot{\rho}_{0g;0e} &=&\left( i\Delta -\frac{\gamma }{2}\right) \rho _{0g;0e}+%
\frac{\eta \Omega }{2}\rho _{0g;1g}\nonumber \\
&&-\varepsilon \left( \rho _{0g;0g}-\rho_{0e;0e}\right) , \\
\dot{\rho}_{0g;1g} &=&\left( i\Delta-\kappa \right) \rho _{0g;1g}-\frac{\eta \Omega }{2}\rho _{0g;0e}+\varepsilon \rho _{0e;1g}, \\
\dot{\rho}_{0e;0e} &=&-\gamma \rho _{0e;0e}+\frac{\eta \Omega }{2}\left( \rho _{0e;1g}+\rho
_{1g;0e}\right) \nonumber \\
&&-\varepsilon \left( \rho _{0g;0e}+\rho _{0e;0g}\right), \\
\dot{\rho}_{0e;1g} &=&-\left( \kappa +\frac{\gamma }{2}%
\right) \rho _{0e;1g}-\varepsilon \rho _{0g;1g} \nonumber \\
&&+\frac{\eta \Omega }{2}\left( \rho _{1g;1g}-\rho_{0e;0e}\right), \\
\dot{\rho}_{1g;1g} &=&-2\kappa \rho _{1g;1g}-\frac{\eta \Omega }{2}\left( \rho _{0e;1g}+\rho
_{1g;0e}\right) .
\end{eqnarray}

We suppose both the internal state and the vibrational mode of the motion of
the ion are initially in their ground state, $\left\vert 0g\right\rangle ,$
that is, $\rho _{0g;0g}(0)=1,\rho _{0e;0e}(0)=\rho _{1g;1g}(0)=\rho
_{0e;1g}(0)=0$. In order to examine the prperties of the refraction and the
absorption for the probe light, we adopt the expression of the complex
susceptibility which is given by $\chi =\chi ^{\prime }+i\chi ^{\prime
\prime }\propto \left( \rho _{0g;0e}/i\varepsilon \right) $ \cite%
{CIT,Scully, Meystre}, where the real part $\chi ^{\prime }$ stands for the
index of refraction of the medium and the imaginary part $\chi ^{\prime
\prime }$ is proportional to the absorption coefficient. Hence our task is
to solve the equations about the elements of the density matrix in order to
obtain $\rho _{0g;0e}$. For simplicity, in the following we only focus on
the steady-state solution of the elements of the density matrix by setting
their first derivatives with respect to time to zero. We finally get the
steady-state solution to $\rho _{0g;0e}$ and hence yielding the information
about the absorption coefficient $\chi ^{\prime \prime }$ and dispersion
coefficient $\chi ^{\prime }$ of a weak probe field. This is given by 
\begin{equation}
\rho _{0g;0e}=\frac{\varepsilon \left( i\Delta -\kappa \right) }{\left(
i\Delta -\frac{\gamma }{2}\right) \left( i\Delta -\kappa \right) +\left( 
\frac{\eta \Omega }{2}\right) ^{2}}.
\end{equation}
The numerical results of $Im[\rho _{0g;0e}/i\varepsilon ]$ are given in Fig. 
$2$, which shows there exists VIT for the probe light when its frequency is
close to the ionic transition frequency. Figure $2(a)$ indicates that the
transparency window becomes wider and deeper as $\Omega $ is getting larger
if the other parameters are unchanged. Figure $2(b)$ shows that as the
heating rate $\kappa$ increases the depth of transparency window becomes
shallow, which indicates the heating effect makes the system more opaque.

\section{ATS in blue-detuning case}

\begin{figure}[bp]
\begin{center}
\includegraphics[
height=2.8in,
width=3.3in,
]{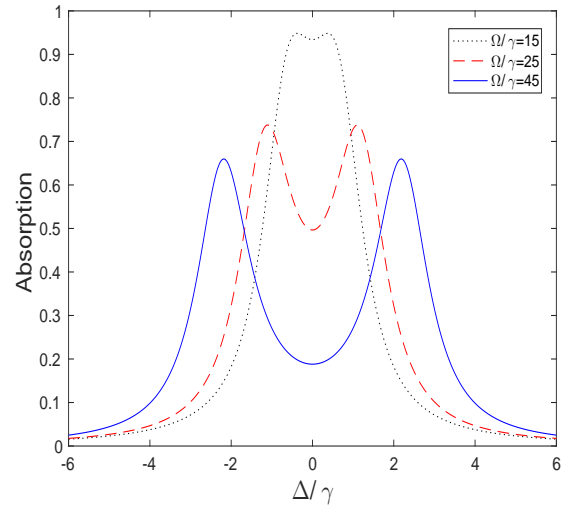}
\end{center}
\caption{The absorption spectra for the probe light when the control light
is tuned to the first blue sideband. The vertical axis is $Im[\protect\rho %
_{0g;0e}/i\protect\varepsilon ]$ in the unit of $1/\protect\gamma $ and the
horizontal axis is $\Delta /\protect\gamma $ with the Lamb-Dicke parameter $%
\protect\eta =0.1$, $\protect\gamma /\protect\kappa =50$ and $\Omega /%
\protect\gamma =15$ (black dotted line), $\Omega /\protect\gamma =25$ (red
dashed line), $\Omega /\protect\gamma =45$ (blue solid line). }
\end{figure}

The master equation for the blue-detuning case can be derived by the way
similar to the red-detuning case and it takes the following form 
\begin{eqnarray}
\dot{\rho} &=&-\frac{i}{\hbar }\left[ H^{\prime \prime },\rho \right]
+\kappa (\overline{n}+1)\left( 2a^{\dagger }\rho a-aa^{\dagger }\rho -\rho
aa^{\dagger }\right)   \nonumber \\
&&+\kappa \overline{n}\left( 2a\rho a^{\dagger }-a^{\dagger }a\rho -\rho
a^{\dagger }a\right) \nonumber \\
&&+\frac{\gamma }{2}\left( 2\sigma _{-}\rho \sigma
_{+}-\sigma _{+}\sigma _{-}\rho -\rho \sigma _{+}\sigma _{-}\right) .
\end{eqnarray}%
Because the ionic vibration is supposed to be confined to its ground state,
the average thermal phonon $\overline{n}$ is almost zero and the motion of
the ion is mostly in the zero- or one-phonon state. According to the master
equation, the elements of the density matrix in the states $\{\left\vert
0g\right\rangle ,$\ $\left\vert 0e\right\rangle ,$\ $\left\vert
1e\right\rangle \}$ are of the form: 
\begin{eqnarray}
\dot{\rho}_{0g;0g} &=&-2\kappa \rho _{0g;0g}-\frac{\eta \Omega }{2}\left( \rho _{0g;1e}+\rho
_{1e;0g}\right) \nonumber \\
&&+\varepsilon \left( \rho _{0e;0g}+\rho _{0g;0e}\right)+\gamma \rho _{0e;0e}, \\
\dot{\rho}_{0g;0e} &=&\left( i\Delta -2\kappa -\frac{\gamma }{2}\right) \rho
_{0g;0e}-\frac{\eta \Omega }{2}\rho _{1e;0e}\nonumber \\
&&-\varepsilon \left( \rho_{0g;0g}-\rho _{0e;0e}\right) , \\
\dot{\rho}_{1e;0g} &=&-\left( 3\kappa +\frac{\gamma }{2}%
\right) \rho _{1e;0g}+\varepsilon \rho _{1e;0e} \nonumber \\
&&+\frac{\eta \Omega }{2}\left( \rho _{0g;0g}-\rho_{1e;1e}\right), \\
\dot{\rho}_{1e;0e} &=&\left( i\Delta -3\kappa -\gamma \right) \rho _{1e;0e}+%
\frac{\eta \Omega }{2}\rho _{0g;0e}-\varepsilon \rho _{1e;0g}, \\
\dot{\rho}_{0e;0e} &=&-\left( 2\kappa +\gamma \right) \rho _{0e;0e}
-\varepsilon \left( \rho _{0g;0e}+\rho _{0e;0g}\right), \\
\dot{\rho}_{1e;1e} &=&-\left( 4\kappa +\gamma \right) \rho
_{1e;1e}+2\kappa \rho _{0e;0e}\nonumber \\
&&+\frac{\eta \Omega }{2}\left( \rho _{0g;1e}+\rho_{1e;0g}\right) .
\end{eqnarray}%
In the same way, the susceptibility is given by $\chi =\chi ^{\prime }+i\chi
^{\prime \prime }\propto \left( \rho _{0g;0e}/i\varepsilon \right) $, $\chi
^{\prime }$ and $\chi ^{\prime \prime }$ are related to the refraction of
the medium and the absorption coefficient, respectively.

As done in Sec. III, the steady-state solution to $\rho _{0g;0e}$ can be
derived by setting the derivatives of the elements of the density matrix in
Eqs. ($20-25$) to zero in the initial condition of $\rho _{0e;0e}(0)=1, \rho
_{0g;0g}(0)= \rho _{1e;1e}(0)=\rho _{1e;0g}(0)=0$. Thus the solution to $%
\rho_{0g;0e}$ is 
\begin{equation}
\rho_{0g;0e}=\frac{\varepsilon \left(i\Delta-3\kappa -\gamma \right)}{%
\left(i\Delta -2\kappa-\frac{\gamma}{2}\right)\left(i\Delta -3\kappa -\gamma
\right)+ \left(\frac{\eta\Omega}{2}\right)^2}.
\end{equation}
In Fig. $3$ we plot the numerical result of $Im[\rho _{0g;0e}/i\varepsilon ]$
as the function of $\Delta/\gamma$. Similar to the red-detuning case, a dip
in absorption spectrum of the probe light emerges slowly in such a case. It
is obvious that the dip becomes deeper and wider with the increase of $\Omega
$ under the condition of the other parameters unchanged.

\section{Discussion and conclusion}

The absorption spectra of the probe light in both cases of red-detuning and
blue-detuning are investigated, and a dip in the spectrum can emerge in both
cases. Differently, in the red-detuning case, the energy level configration
is of $\Lambda $-type three structure and the dip in absorption spectrum
exhibits the properties of EIT, that is, a narrow and deep dip can appear
when the driving light is not so strong, while in the blue-detuning case,
the energy level configration takes $V$-type three structure and the dip
exhibits the properties of ATS, the appearance of the dip requires a
stronger driving light and the dip is either narrow but shallow or deep but
wide, i.e., the dip cannot be narrow and deep at the same situation.

Our proposal about the VIT may be verified experimentally. On the one hand,
the techniques for ion traps have been utilized to realize much complicated
quantum process \cite{RMP-ion} and quantum logic gates \cite%
{Cirac-ZollerEXP,iontrapEXP,Wineland2,Wineland1} as mentioned in Sec. I,
such techniques pave a way for the VIT and ATS presented here. On the other
hand, vaccum induced transparency in a cavity \cite{VaIT} indicates that the
transparency of light can be achieved for several or even a single atom(s),
thus our proposal should be realized experimentally.

To summarize, in the present work we have investigated ionic vibration
induced transparency and Autler-Townes splitting in a linear Paul trap. When
control light is tuned to the first red sideband of the ionic transition,
the VIT emerges and it is very similar to the CIT \cite{CIT}. When the
control light is tuned to the first blue sideband of the ionic transition,
the ATS emerges via anti-JC Hamiltonian. We find in both cases the dip in
the absorption spectra becomes wider and deeper as the Rabi frequency of the
control light increases.

\section*{Acknowledgement}

This work is supported by the Natural Science Foundation of Shanghai 
(Grant No. 15ZR1430600), National Natural Science Foundation of China  
under Grant Nos. 61475168, 11674231, 11574179 and 11074079. XLF is sponsored 
by Shanghai Gaofeng \& Gaoyuan Project for University Academic Program Development.

\end{document}